\documentclass{PoS}

\usepackage{nicefrac}

\title{Direct photon emission in Heavy Ion Collisions from Microscopic
Transport Theory and Fluid Dynamics
}

\ShortTitle{Direct Photons In Heavy Ion Collisions}

\author{\speaker{Bj\o{}rn~B\"auchle}~and
        Marcus~Bleicher\\
        Frankfurt Institute for Advanced Studies, Frankfurt am Main,
        Germany\\
        Institut f\"ur Theoretische Physik Frankfurt am Main, Germany \\
        E-mail: \email{baeuchle@th.physik.uni-frankfurt.de},
        \email{bleicher@th.physik.uni-frankfurt.de}}

\abstract{
Direct photon emission in heavy-ion collisions is calculated within a
relativistic micro+macro hybrid model and compared to the microscopic
transport model UrQMD.  In the hybrid approach, the high-density part of the
collision is calculated by an ideal 3+1-dimensional hydrodynamic
calculation, while the early (pre-equilibrium-) and late (rescattering-)
phase are calculated with the transport model. Different scenarios of the
transition from the macroscopic description to the transport model
description and their effects are studied. The calculations are compared to
measurements by the WA98-collaboration and predictions for the future
CBM-experiment are made.
}

\FullConference{XLVIII International Winter Meeting on Nuclear Physics,
                BORMIO2010\\
		January 25-29, 2010\\
		Bormio, Italy}

\begin{document}

\section{Introduction}

Electromagnetic Probes provide a unique insight into the early stages of
heavy-ion collisions, since they have the advantage of negligible
rescattering cross-sections. Therefore, they leave the production region
without rescattering and carry the information from this point to the
detector. Besides single- and dileptons, direct photon emission can
therefore be used to study the early hot and dense, possibly partonic,
stages of the reaction.

Unfortunately, most photons measured in heavy-ion collisions come from
hadronic decays. The experimental challenge of obtaining spectra of only
direct photons has been gone through by several experiments; WA98
(CERN-SPS)~\cite{Aggarwal:2000ps} and PHENIX
(BNL-RHIC)~\cite{Adler:2005ig} have published explicit data points
for direct photons.

On the theory side, the elementary photon production cross-sections are
known since long, see e.g.\ Kapusta {\it et al.}~\cite{Kapusta:1991qp} and
Xiong {\it et al.}~\cite{Xiong:1992ui}.  The major problem is the difficulty
to describe the time evolution of the produced matter, for which first
principle calculations from Quantum Chromodynamics (QCD) cannot be done.
Well-developed dynamical models are therefore needed to describe the
space-time evolution of nuclear interactions. 

Among the approaches used are relativistic transport
theory~\cite{Geiger:1997pf,Bass:1998ca} and relativisitc fluid- or
hydrodynamics~\cite{Cleymans:1985wp,Rischke:1995mt}.  For both models,
approximations have to be made, and in both models, the restrictions imposed
by these approximations can be loosened. For transport theory, the necessary
approximations include the restriction of scattering processes to two
incoming particles, which limits the applicability to low particle
densities. For hydrodynamics, on the other hand, matter has to be in local
thermal equilibrium (for ideal, non-viscous hydrodynamic calculations) or at
least close to it (for viscous calculations).

From these deliberations, it is clear where the advantages for both models
are: While in transport, non-equilibrium matter, which is present in the
beginning of the heavy-ion reaction, and dilute matter, which is present in
the late phase, can be described, hydrodynamics may be better suited to
describe the intermediate stage, which is supposed to be dense, hot and
thermalized. In addition, the transition between two phases of matter, such
as Quark Gluon Plasma (QGP) and Hadron Gas (HG) can be easily described in
hydrodynamics, while this is not (yet) possible for transport models, since
the microscopic details of this transition are not known.

\section{The Model}\label{sec:model}

UrQMD v2.3 (Ultra-relativistic Quantum Molecular Dynamics) is a microscopic
transport model~\cite{Bass:1998ca,Petersen:2008kb}. It
includes all hadrons and resonances up to masses $m \approx 2.2~{\rm GeV}$
and at high energies can excite and fragment strings. The cross-sections are
either parametrized, calculated via detailed balance or taken from the
additive quark model (AQM), if no experimental values are available. In the
UrQMD framework, propagation and spectral functions are calculated as in
vacuum.

In the following, we compare results from this microscopic model to results
obtained with a hybrid model description~\cite{Petersen:2008dd}. Here, the
high-density part of the reaction is modelled using ideal 3+1-dimensional
fluid-dynamics.  The unequilibrated initial state and the low-density final
state are described by UrQMD. Thus, those stages have only hadronic and
string degrees of freedom.

To connect the initial transport phase with the high-density fluid phase,
the baryon-number-, energy- and momentum-densities are smoothed and put into
the hydrodynamic calculation after the incoming nuclei have passed through
each other. Temperature, chemical potential, pressure and other macroscopic
quantities are determined from the densities by the Equation of State used
in the current calculation. During this transition, the system is forced
into an equilibrated state.

In non-central collisions, the spectators are propagated in the cascade.
After the local rest frame energy density has dropped below a threshold
value of $\epsilon_{\rm crit} \approx 5 \epsilon_0$, particles are created
on a hyper-surface from the densities by means of the Cooper-Frye formula
and propagation is continued in UrQMD.

The transition scenario used in the calculations presented here can be
either isochronous, i.e.\ all particles are created at the same time, or
``gradual''. In the latter scenario, particles are created at the same time
for every slice in the longitudinal direction. This represents a
pseudo-eigentime condition.

For these investigations, we use three different Equations of State for the
hydrodynamic phase. The base line calculations are done with a Hadron Gas
Equation of State (HG-EoS), which includes the same degrees of freedom as
present in the transport phase. This allows to explore the effects due to
the change of the kinetic description. Secondly, we use a MIT-Bag Model EoS
(BM-EoS) with a partonic phase and a first order phase
transition~\cite{Rischke:1995mt}. We use the BM-EoS for investigations of
photon emission from the QGP. The third Equation of State $\chi$-EoS used
here has a chirally restored phase with a critical end
point~\cite{Steinheimer:2009hd}.

Photon emission is calculated perturbatively in both models, hydrodynamics
and transport, because the evolution of the underlying event is not altered
by the emission of photons due to their very small emission probability. The
channels considered for photon emission may differ between the hybrid
approach and the binary scattering model.  Emission from a
Quark-Gluon-Plasma can only happen in the hydrodynamic phase, and only if
the equation of state used has partonic or chirally restored degrees of
freedom. Photons from baryonic interactions are neglected in the present
calculation.

For emission from the transport part of the model, we use the
well-established cross-sections from Kapusta {\it et
al.}~\cite{Kapusta:1991qp}, and for emission from the hydrodynamic phase, we
use the parametrizations by Turbide, Rapp and Gale and Arnold {\it et
al.}~\cite{Turbide:2003si} (the latter for QGP-emission).  For detailed
information on the emission process, the reader is referred to B\"auchle
{\it et al.}~\cite{arXiv:0905.4678}.

\section{Results}

\begin{figure}
 \includegraphics[width=\textwidth]{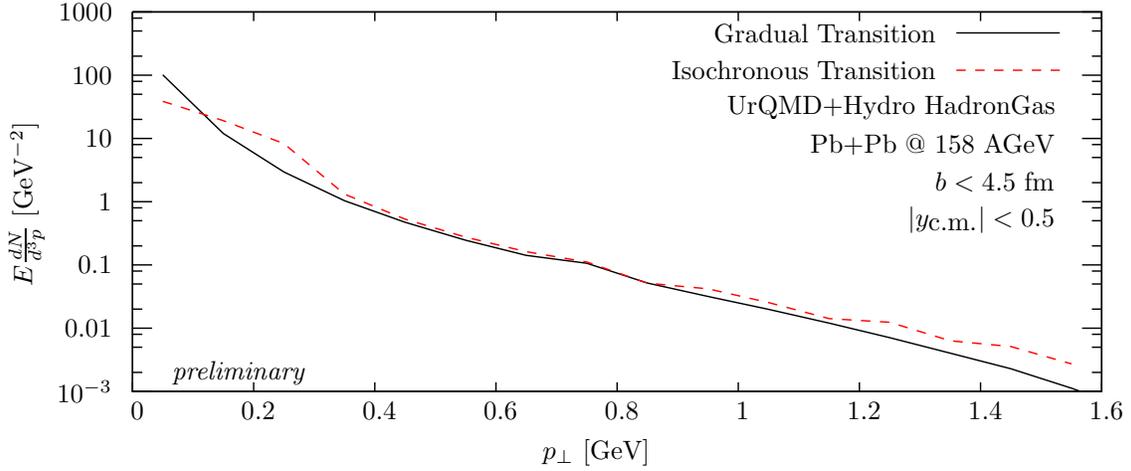}
 \caption{Direct photon spectra from hybrid-model calculations with hadron
 gas EoS. The calculation with isochronous transition is shown as a red
 dotted line, the calculation with gradual transition as black solid line.}
 \label{fig:gradual}
\end{figure}

\begin{figure}
 \includegraphics[width=\textwidth]{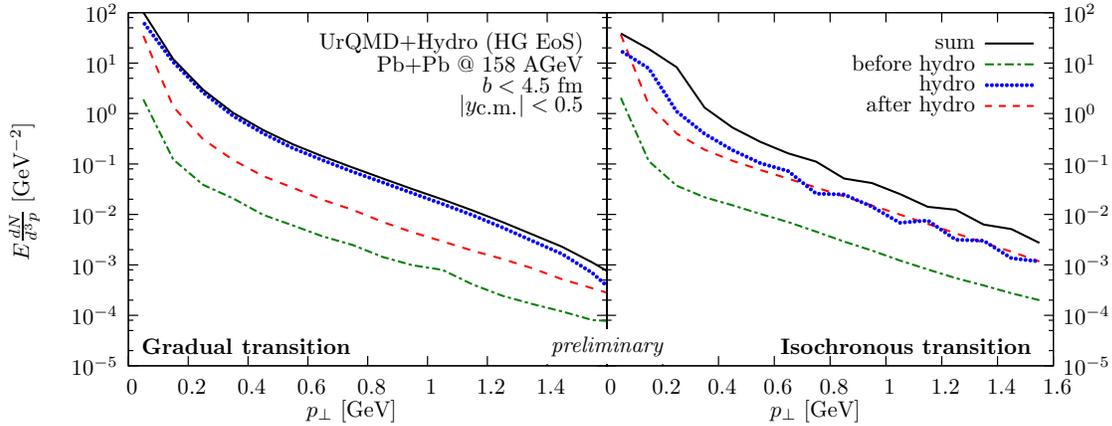}
 \caption{Contribution of the stages before (dark green dash-dotted),
 during (blue dotted) and after (red dashed) hydrodynamic
 evolution with gradual transition (left-hand side) and isochronous
 transition (right hand side).}
 \label{fig:stages}
\end{figure}

In Figure~\ref{fig:gradual}, we compare inclusive spectra from hybrid
calculations with isochronous and gradual transition from the hydrodynamics
to cascade phases. Here, we find the spectra to be very consistent with each
other. But when looking at the contributions of the different stages -- the
early, intermediate (hydrodynamic) and late stage -- in
Figure~\ref{fig:stages}, we find significant differences in the relative
contributions of intermediate and late stage.  While for the isochronous
transition, both phases contribute in similar amounts, the gradual
transition scenario is dominated by the hydrodynamic intermediate stage and
has a greatly reduced late stage emission.

\begin{figure}
 \includegraphics[width=\textwidth]{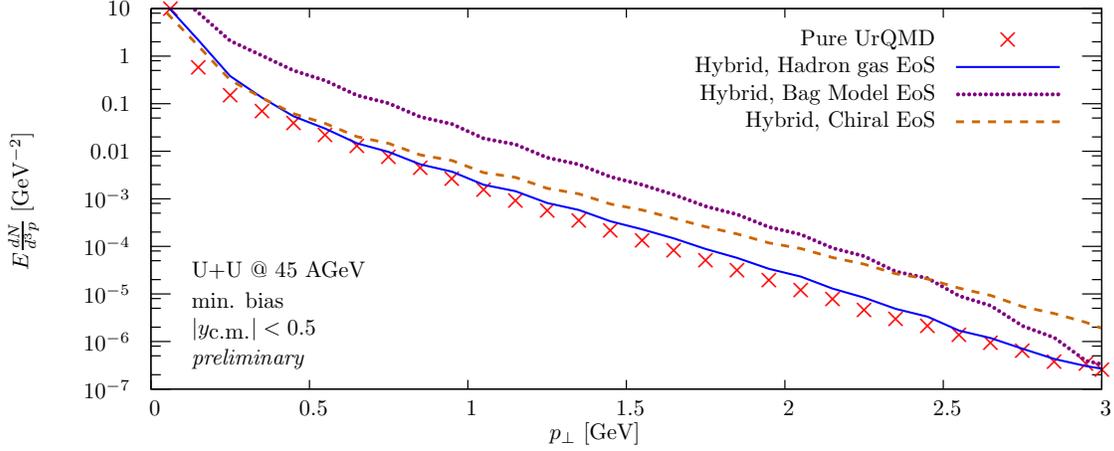}
 \caption{Comparison of the direct photon spectra with different variations
 of the model for FAIR-CBM-energies. The calculations are done with the
 ``isochronous'' transition scenario (see text). We show calculations
 without hydrodynamic state (red crosses), hadron gas EoS (blue solid), Bag
 model EoS (violett dotted) and chiral EoS (orange dashed line).}
 \label{fig:fair}
\end{figure}

Photon emission has also been calculated for minimum bias $U+U$-collisions
at $E_{\rm lab} = 45$~AGeV, as are planned at the future FAIR-facility (see
Figure~\ref{fig:fair}). At
these energies, the pQCD-contribution from proton-proton collisions is
negligible. We can confirm that the cascade-calculation and the hybrid
calculation with Hadron Gas EoS yield consistent results, as was found in
calculations for $E_{\rm lab} = 158$~AGeV (see~\cite{arXiv:0905.4678}). The
emission in Bag Model EoS calculations is greatly enhanced with respect to
the hadronic base line calculations, and the results from using the Chiral
EoS show a significant enhancement at large $p_\bot$. Thus, experiments at
FAIR will be well-suited to distinguish the different models.

\section{Summary}

With UrQMD, we have explored the impact of the transition scenario from the
hydrodynamic to the cascade phase on the emission of photons. The main
result is that with the standard model parameters, the effect of changing
this transition is negligible. A closer look at the origin of the photons
and the contribution of the different stages suggests that the comparison
between gradual and isochronous transition scenarios is likely to depend on
the criterium for this transition -- i.e. at what energy density this
transition happens.

At low beam energies, where pQCD-effects do not contribute to the overall
spectra, a clearer comparison between different scenarios will be possible.
Specifically, calculations with different EoS yield significant differences.

\section{Outlook}

The results shown here suggest the need for further studies. Especially the
different contributions of the stages between the transition scenarios
suggest a closer inspection of the transition parameters and its impact on
photon emission. The time of the first transition from cascade to
hydrodynamics should also be examined. All of these studies will be carried
out for various equations of state.

The preliminary results for the FAIR-system $U+U$ at $E_{\rm lab} = 45$~AGeV
show significant differences between the various Equations of State, which
will be examined in more detail in the near future.

\acknowledgments

This work has been supported by the Frankfurt Center for Scientific
Computing (CSC), the GSI and the BMBF. The authors thank Hannah Petersen for
providing the hybrid- and Dirk Rischke for the hydrodynamic code. B.\
B\"auchle gratefully acknowledges support from the Deutsche Telekom
Stiftung, the Helmholtz Research School on Quark Matter Studies and the
Helmholtz Graduate School for Hadron and Ion Research. This work was
supported by the Hessian LOEWE initiative through the Helmholtz
International Center for FAIR.

The authors thank Elvira Santini, Pasi Huovinen and Rene Bellwied for
valuable discussions and Klaus Reygers for experimental clarifications.

\end{document}